\documentclass[%
 reprint,
nofootinbib,
 amsmath,amssymb,
 aps,
pra,
]{revtex4-1}

\usepackage{braket}
\usepackage{amsmath}
\usepackage{amsthm}
\usepackage{amssymb}
\usepackage{amsfonts}
\usepackage{mathtools} 
\usepackage{graphicx}
\graphicspath{{images/}}
\usepackage{dcolumn}
\usepackage{bm}

\usepackage{float} 
\usepackage{lipsum} 

\usepackage{mathpazo} 
\usepackage{avant} 

\usepackage[para]{threeparttable}
\usepackage{array,booktabs,longtable,tabularx,tabulary}
\newcolumntype{L}{>{\raggedright\arraybackslash}X}
\usepackage{ltablex}
\usepackage{siunitx}
\usepackage[flushleft]{threeparttablex}
\usepackage{array}
\newcolumntype{P}[1]{>{\centering\arraybackslash}p{#1}}

\usepackage[caption=false]{subfig}

\newtheorem{lemma}{Lemma}
\newtheorem{prop}{Proposition}
\newtheorem{defi}{Definition}

\newtheorem{innercustomdefi}{Proposition}
\newenvironment{customdefi}[1]
  {\renewcommand\theinnercustomdefi{#1}\innercustomdefi}
  {\endinnercustomdefi}

\usepackage{chngcntr}
\newcommand{\tr}[1]{\operatorname{tr}(#1)}

\usepackage{hyperref}
\usepackage{cleveref} 

\hypersetup{
    colorlinks=true,
    linkcolor=red,
    filecolor=magenta,      
    urlcolor=cyan,
}
\usepackage{xcolor}
\renewcommand\theequation{{\color{blue}\arabic{equation}}}

\usepackage{appendix}

\usepackage{dashbox}

\begin{document}

\title{Quantifying quantum reference frames in composed systems: local, global and mutual asymmetries}

\author{Tiago Martinelli}%
\email{tiago.martinelli93@gmail.com}
\affiliation{Instituto de F\'isica de S\~ao Carlos, Universidade de S\~ao Paulo, CP 369, 13560-970, S\~ao Carlos, SP, Brazil}

\author{Diogo O. Soares-Pinto}
\email{dosp@usp.br}
\affiliation{Instituto de F\'isica de S\~ao Carlos, Universidade de S\~ao Paulo, CP 369, 13560-970, S\~ao Carlos, SP, Brazil}

\date{\today}

\begin{abstract}
The Page-Wootters mechanism questioned the fundamental nature of time in quantum physics. The mechanism explored the notion that a given physical quantity is always defined and measured relative to a reference frame, in general, not explained in the theoretical description of quantum physical experiments. Recently, the resource theory of asymmetry deals explicitly with what are the physical conditions for a quantum system to serve as a good reference frame. Nonetheless, to quantify a quantum reference frame in relation to another one it is a important task to establish an internal description of quantum theory, i.e., without the need of a classical reference frame. In this work we address this issue by the concept of mutual asymmetry and use this machinery in the Page-Wootters mechanism by identifying the concept of mutual asymmetry as mutual or internal coherence. To do so, the notion of quantum coherence in relation of a quantum reference frame is revisited and a quantifier is proposed in this scenario. Also, this open space to investigate the link of internal coherence and correlations, as proposed by Page and Wootters, under a resource theory approach.
\end{abstract}

\maketitle

\section{\label{sec:level1}Introduction}

The ordinary practice of science describe systems of the universe against a background reference frame, idealized to be fixed. By Smolin \cite{smolin1}, it is argued that the concept of time is the key ingredient when one consider including the whole universe within one's system. Then, all observables are described in relation to dynamical reference frames without the need of an external, absolute reference frame. In this view, time is supposed to be emergent from a timeless fundamental theory \cite{smolin2}. A timeless picture of nature have pre-Socratic Greek roots and resides on the classical debate between Parmenides and Heraclitus \cite{parmenides}.\newline

In this paper, following Smolin lines we aim at exploring an internal description of quantum theory. This implies to investigate a timeless approach by the role of quantum reference frames inside the theory. To do so, we could start by using the operational definition of time in quantum mechanics given by Peres: ''time is what is measured by a clock'' \cite{macone}. This statement could solve the problem of defining time, but in fact it just reshapes the question to say: what indeed are clocks inside quantum theory?\newline

The idea behind the current-technological clocks is to use an atomic transition in a known frequency/energy to calibrate a single mode laser such that its frequency is stable at the atomic frequency. Since the transition frequency can be measured with a high degree of accuracy, it provides a standard frequency of the laser \cite{muga}. Even though it uses a quantum mechanism, the time of the clock - represented by the frequency source of the laser here, is kept by a digital counter which counts the number of elapsed periods. This extra apparatus is classical and it is not subject to the laws of quantum mechanics, therefore the total system does not provide a fully quantum clock \cite{pashby}. By making a fully quantum analysis in the mechanism \textit{plus} apparatus it could raise a way of modelling a quantum observer \cite{,faist}. Besides, to investigate the relational approach to quantum theory, which suggests that features of a system such as entanglement and superposition are observer-dependent \cite{brukner_intro}. Problems of this kind appear when the interface classical-quantum reference frames are investigated in the quantum theory \cite{love1}. \newline

If one tries to define a quantum clock as a system $H_{R}$ with a Hilbert space structure in which the eigenvalues of the clock operator system $T_{R}$ satisfies $[T_{R},\,H_{R}]=i\mathbb{1}_{R}$ and gives the elapsed parametric time, then this whole mechanism is failed by the Pauli's argument \cite{02,58p, pashby}. To circumvent this, a possible path consists in the idea to include the quantum clock in an extend Hilbert space  and consider the composed space as the proper physical working place \cite{15}. Historically, the extension of Hilbert space is due to Dirac \cite{pashby}, and later he used the formalism motivated by the desire to quantify the general relativity \cite{64}. Such techniques was also used by John Wheeler and Bryce DeWitt on their time-static equation in a desire to obtain an theory for quantum gravity \cite{67}. Finally, such proposal was one of the Page and Wootters motivations \cite{83,84} for the mechanism which we are going to discuss here. Recently, the interest by this mechanism is being revisited in literature \cite{14m,15,16,ahmadi,rossignoli1,rossignoli2,new1}.\newline

In this paper we show that in order to give a internal description of a system relative to a quantum reference frame, both inside a globally symmetric composed system $\mathcal{H}$, the resource needed is the existence of \textit{mutual asymmetry}. We illustrate this concept by considering our internal quantum reference frame as a quantum clock. In this particular case, the mechanism in question is the Page-Wootters clock (PWC) and the resource turns to be the existence of \textit{mutual or internal coherence} \cite{new2} between the system and the clock. For this purpose, we provide an asymmetry quantifier capable to deal with composed systems to describe the physical phenomena as well as its operational meaning and regime analysis.\newline

In what follows, we give some necessary preliminaries before stating our results: in Sec.\ref{sec:level2} we introduce the concept of quantum reference frames from quantum clocks in the PWC model, and in Sec.\ref{sec:level3} we introduce some tools from resource theory of translational-asymmetry or quantum reference frames. Finally, in Sec.\ref{sec:level4} we exhibits our results giving an operational formalism for the role of reference frames inside composed system. We apply to illustrative examples treating time-asymmetry as shifts in phase related by unitary representations of U(1) group.


\section{\label{sec:level2}Quantum reference frames in the PWC model}

In non-relativistic quantum theory, a symmetry group $G$ acts in the Hilbert space $\mathcal{H}$ of a given system via a (strongly continuous, projective) unitary representation $U$, in this case $G$ is known as the Galilei group \cite{love1}. By simplicity, in this paper we will deal only with unitary representations promoting translations in one dimension exemplified as phases given by the U(1) group, despite the generalization for other cases can be treated.\newline

The PWC model \cite{83,84,15,16} argued that the notion of time appears from correlations between a system $S$ and a reference system $R$ in a composed system under global time-symmetry. Such composition consists in an extension of the Hilbert space $\mathcal{H}_{S}$ to
$\mathcal{H}\coloneqq\mathcal{H}_{S}\otimes\mathcal{H}_{R}$, where $\mathcal{H}_{R}$ is the clock reference space. The global time-symmetry imposes that the total Hamiltonian of the system $H\coloneqq H_{S}\otimes \mathbb{1}_{R}+\mathbb{1}_{S}\otimes H_{R}$, with $\mathbb{1}_{\alpha}$ the identity operator in system $\alpha = S,R$,  satisfies,
\begin{equation}
\left.H\left.|\psi\right\rangle \right\rangle =0,
\end{equation}in which the double-ket notation means that $\left.\left.|\psi\right\rangle \right\rangle \in \mathcal{H}=\mathcal{H}_{S}\otimes\mathcal{H}_{R}$. In the density operator formalism, this condition can be written as:
\begin{equation}
    [H,\rho] = 0, 
\label{first}
\end{equation}with $H,\rho\in \mathcal{H}_{S}\otimes\mathcal{H}_{R}$. The mechanism itself codifies the external time (imposes global time-symmetry), a non-observable quantity, by the technique of time averaging \cite{89}:

\begin{equation}
\mathcal{G}(\rho)=\intop_{g\in G}d\mu(g)\,\mathcal{U}_{g}^{SR}(\rho),
\
\label{eq:twirling}
\end{equation}with $\mathcal{U}_{g}^{SR}(\cdot)=U_{g}^{S}\otimes U_{g}^{R}(\cdot)U_{g}^{S\dagger}\otimes U_{g}^{R\dagger}$
acting in the  system $S+R$. This operation is the uniform twirling over a given group $G$, which transforms its input to a symmetric state \cite{bartlett}. The integral exists only for groups with well-defined Haar measure $d\mu(g)$ \cite{haar} (for the U(1)-group the measure is $\frac{d\theta}{2\pi}$). This encoding map means physically that given a quantum system $S$, one wants to introduce a quantum reference frame - the clock space $R$, to give a fully quantum description of time.\newline

The  clock reference space $\mathcal{H_{R}}$ breaks time-symmetry by indicating the pointer orientation $g$ associated with the time-symmetry group $G$ generated by ${H}_{R}$. Following \cite{palmer1}, one could construct the set of clock states by starting with the state $\ket{\phi_{R}(e)}$, which serves as the zero time oriented with respect to a background frame and is associated with the identity $e \in G$. To construct states corresponding to other orientations $g\in G$ one generates the states in the orbit of $\ket{\phi_{R}(e)}$ under the group action $U_{g}^{R}=e^{-iH_{R}g}$ (throughout this paper we assume $\hbar \equiv 1$), giving $\ket{\phi_{R}(g)} = U_{g}^{R}\ket{\phi_{R}(e)}$, $\forall g\in G$. These states satisfies $U_{h}\ket{\phi_{R}(g)} = \ket{\phi_{R}(hg)},\, h,g\in G$, which means that they transform covariantly under the action of the time-symmetry group. Ideally, there exists a self-adjoint clock operators $T_{R}$, such that $[H_{R},\,T_{R}]=i\mathbb{1}_{R}$ \cite{15}. This guarantees the generator $H_{R}$ promoting shifts in the clock operator, $U_{\Delta g}^{R}T_{R}U_{\Delta g}^{R\dagger}=T_{R}+\Delta g\mathbb{1}_{R}$, giving the distinguishable basis of time states $\{\ket{\phi_{R}(g)}\}_{g\in G}$ as defined above. \newline

It is worth to mention that the conditions above promotes the PWC model to recover the Schr{\"o}dinger dynamics for the system $S$ as well as the formalism of conditional probability to measurements in quantum theory. However, our focus on this paper is to investigate the physics of quantum reference frames in the PWC, for a brief review of these extra conditions, see appendix \ref{appA} and Refs.\cite{time,monte1}.

\subsection*{\label{sub_qbits}A qubit as quantum reference frame} 

Before proceeding, we can see an example of the smaller possible case of quantum reference frame and its consequences - the qubit model worked in Ref.\cite{83}. The total Hamiltonian is: 
\begin{equation}
H=\sigma_{z}^{R}\otimes I^{S}+I^{R}\otimes \sigma_{z}^{S},
\end{equation}with $\sigma_{z}^{\alpha}$ being the Pauli operator in $\hat{z}$-direction, of each particle. 

To make the clock space clearer here, one could take the following picturesque assignment of ''hours'' to the states of the quantum clock as follows \cite{84}: to the state corresponding to $+\hat{x},$ assign $\left.| + \right \rangle\longleftrightarrow 12h$, and to the state corresponding to $-\hat{x},$ assign $ \left. |- \right \rangle \longleftrightarrow 6h$, representing a lag angle of $\pi$. In other words, to have distinguishability, $\left.|\phi_{12}\right\rangle \equiv\frac{\left.|0\right\rangle +\left.|1\right\rangle }{\sqrt{2}}$ and $\left.|\phi_{6}\right\rangle \equiv\frac{\left.|0\right\rangle -\left.|1\right\rangle }{\sqrt{2}}$, implying $T_{R}\equiv\sigma_{x}^{R}$. Therefore, we have that,

\begin{equation}
e^{-i\sigma_{z}\frac{\pi}{2}}\left.|\phi_{12}\right\rangle=\left.|\phi_{6}\right\rangle,
\end{equation} and the clock is a binary of tics: up-down. This implies that $[\sigma_{z}^{R},\,\sigma_{x}^{R}]=i\sigma_{y}^{R}\neq i\mathbb{1}_{R}=[H_{R},\,T_{R}]$, due the discrete character of the clock. Furthermore, there is an uncertainty in the orientations $+\hat{x},$ and $-\hat{x},$ given by the variance of $\ket{\phi_{12}}$ and $\ket{\phi_{6}}$, respectively, due the fact that there are only two eigenstates in the clock system $R$ to assign hours. Therefore, the closer the eigenvalues of the clock system approach the real line, the better the chance of assigning more time intervals with smaller variance to the dynamics of the system \cite{90d}. Indeed, we will see quantitatively in Sec.\ref{sec:level4} that for the construction to be compatible with a realistic dynamics one must have a high degree of degenerescence in the eigenspace associated with the null eigenvalue of the Hamiltonian of the total system due the high dimension of the clock reference space.\newline

\section{\label{sec:level3}Quantum reference frames as resources for asymmetry}

To approximate the ideal commutator relationship between a time operator and Hamiltonian one needs a  continuous evolution of the clock states \cite{80}. Beyond that, to guarantee quantum features when building the quantum reference clocks it is important to use limited finite resources \cite{new2}. Therefore we will use a finite continuous quantum clock as model to be detailed in appendix \ref{appB}. To impose symmetry in the systems, we will make use of compact Lie groups.\newline

When modelling a dynamics by finite-dimensional representations $U_{g}$ of a continuous Lie group $G$, the reference states $\ket{\phi_{R}(g)}$ for different orientations cannot be perfectly distinguishable which promotes an uncertainty in the orientation $g$, \cite{new1}. One way to quantify these finite resources is to use the dimensionality of the Hilbert space $H_{R}$, which can be constrained by the number of charge sectors $k_{R}$ under the representation of the group in question \cite{palmer1}. To have a well-defined classical limit, we imposes that the overlap of the reference states with different orientations becomes zero as the size parameter $k_{R}$ increases to infinity, 
\begin{equation}
    \lim_{k_{R}\rightarrow\infty}d_{k_{R}}|\braket{\phi_{R}(g)|\phi_{R}(h)}|=\delta(g h^{-1}),
\end{equation}with $\delta(g)$  the delta function on $G$ \cite{palmer1}, and $d_{k_{R}}$ the dimension of $H_{R}$ spanned by $\{\ket{\phi_{R}(g)};g\in G\}$.\newline

To maximize the distinguishability of the quantum reference frame in the finite size case according with equation above would be interesting it scales with $d_{k_{R}}$. A possible choice of reference states for attending this purpose are the maximum likelihood states \cite{chiribella}. In the case of PWC model, the clock states are built from uniform superpositions in the energy eigenstates of $H_{R}$, and so are maximally coherent in energy \cite{new1}. Therefore, if one's interest is to deal with quantum reference frames for time, i.e., resources in the context of asymmetry relative to the group of time translations, these can be understood as dealing with quantum clocks by using the resource of coherence.\newline

\subsection*{\label{subB}Coherence as resource for time-asymmetry}

To deal with resource theories of coherence it is worth to mention that in recent years, it has been established two slightly different approaches: the first approach, due to Baumgratz et al. \cite{baum} and \text{\AA}berg's \cite{aberg,aberg2, gerardo} is aimed at developed a coherence quantifier and its set of conditions which must to be fulfilled, \cite{gerardo}. In the second approach, the resource theory for quantum coherence is viewed as a particular case of the more general theory of asymmetry \cite{jacobs,marvian3,marvian5, gour-spekkens}. On the later, coherent states can serve as resources to overcome the conservation laws in the presence of a given symmetry \cite{susskind,wick,kitaev,bartlett}. For the PWC model,  the set of free states on $\mathcal{H}_{S}\otimes \mathcal{H}_{R}$ are defined as the states which are invariant under all time translations. Similarly, free operations $\mathcal{E}$ on $\mathcal{H}_{S}\otimes \mathcal{H}_{R}$ is defined as a completely positive trace-preserving map (CPTP) invariant under all time translations, satisfying the requisites of a resource theory \cite{andreas}. In graphical words, imposes the following arrow together with the commutative diagram,

\begin{figure}[H]
\centering

\includegraphics[width=6cm]{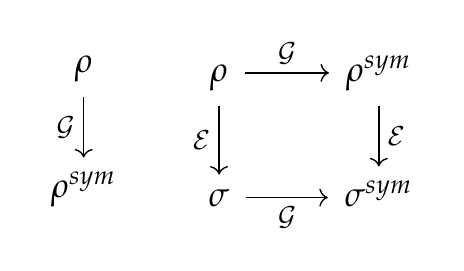}
\caption{Free operations in relation to the group translations, where $\mathcal{E}$ $G$-invariant, and $\mathcal{G}(\cdot) = \intop_{g\in G}d\mu(g)\,\mathcal{U}_{g}^{SR}(\cdot)$ with $\mathcal{U}_{g}^{SR}(\cdot)=U_{g}^{S}\otimes U_{g}^{R}(\cdot)U_{g}^{S\dagger}\otimes U_{g}^{R\dagger}$
acting in the $S+R$ system.}

\end{figure}

In any resource theory, to make the resources useful it is important to be able to quantify them. This is the role of monotones or measures of the resource. Following Ref. \cite{marvian1}, we have that,

\begin{equation}
A_{G}(\rho)\coloneqq S(\mathcal{G}(\rho))-S(\rho),\label{eq:ent_assym}
\end{equation}named relative entropy of asymmetry, defines a measure of asymmetry for states in relation to translational symmetry \cite{marvian3}. This same function has also been studied by \text{\AA}berg \cite{aberg2} under the name of relative entropy of superposition for the particular case of time-translational symmetry. In this case, the uniform twirling turns to be the dephasing map, i.e., the map that dephases its input relative to the eigenbasis of the energy.\newline

The fact that the resource of time-asymmetry or quantum coherence is only defined relative to a choice of basis raises the relational character for time as argued by Page and Wootters. For a more general view, this raises a need for a relational understanding for the resource theory of asymmetry, capable to clarify the role of the standard classical reference system $C$ from quantum reference frames $R$. Furthermore, how should the relative entropy of asymmetry be formulated in such a way capable to distinguish simple from composed systems?

\section{\label{sec:level4}Results and discussion}

To start, assume the presence of a classical reference frame $C$ recording a classical information represented here by a group element $g$, via $g\rightarrow \ket{g}\bra{g}^{C}$, with $\{\ket{g} ;\, g \in G\}$ an orthogonal set of states spanning the Hilbert space of $C$. The whole universe system can be described by the classical-quantum state (cq-state) \cite{marvian2}:

\begin{equation}
\Omega_{CSR}=\intop_{g\in G}d\mu(g)\,\ket{g}\bra{g}^{C}\otimes\mathcal{U}_{g}^{SR}(\rho_{SR}).
\label{omega1}
\end{equation}

\begin{figure*}

\centering
\includegraphics[width=13cm,height=7cm]{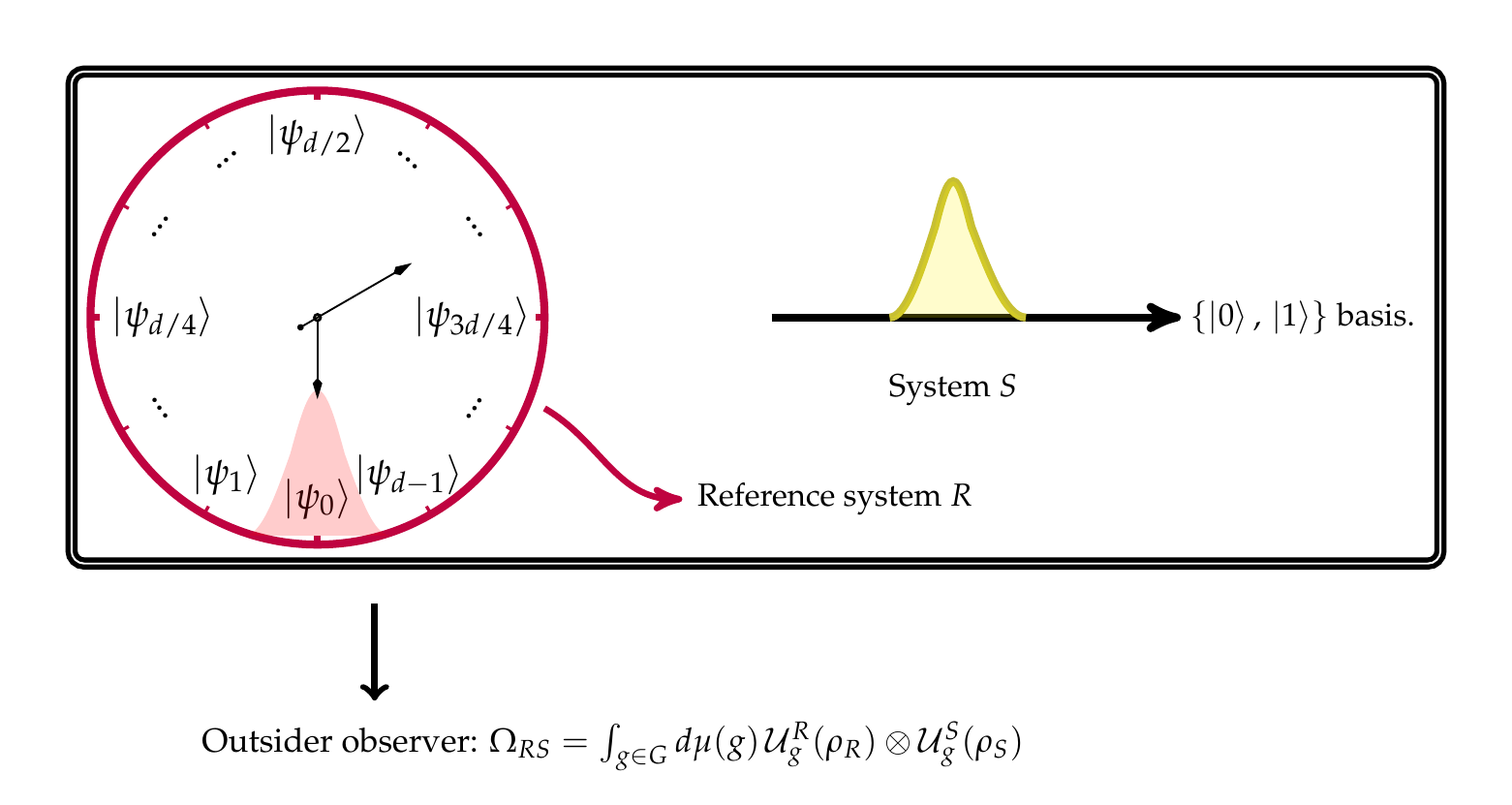}

\caption{In the case where $\rho_{SR} = \rho_{S}\otimes \rho_{R}$, with $\rho_{S}=\ket{\psi_{S}}\bra{\psi_{S}}$, $\left.|\psi_{S}\right\rangle =\frac{\left.|0\right\rangle +\left.|1\right\rangle }{\sqrt{2}}$ and $\rho_{R}=\ket{\psi_{R}}\bra{\psi_{R}}$, $\left.|\psi_{R}\right\rangle =\frac{1}{\sqrt{d}}\sum_{m=0}^{d-1}\ket{m}$ this could describe the PWC model, where the quantum reference clock $\rho_{R}$ can be treated as a quantum observer (QO) \cite{faist}. The preparation of coherent superpositions of energy levels on $S$ and $R$ is discussed along the text. Now, the whole system $\Omega_{SR}$ has a internal dynamics described by the QO observing the system $S$. The supposed outsider observer explain this by a correlated and symmetric state between the system and the quantum clock.}
\label{pwc_figure}
\end{figure*}

In the case of th PWC model, this could be how one orienties the quantum composed system $S+R$ (already prepared previously) in relation to the classical clock $C$. A measurement in the $\ket{g}$ basis on $C$ provides the state $\rho_{SR}$ at instant $g$. Therefore, to make a fully quantum analysis, we have to consider the state $\Omega_{SR}=\mbox{tr}_{C}\Omega_{CSR}$\footnote{Here, to avoid a formalism of the partial trace and measurements in continuous distributions we are making use of the lemma \ref{lemma_approx} in the appendix \ref{appC}.} as the whole universe now, see Fig. \ref{pwc_figure}:

\begin{equation}
\Omega_{SR}=\intop_{g\in G}d\mu(g)\,\mathcal{U}_{g}^{SR}(\rho_{SR}).
\label{omega2}
\end{equation}

To give a general quantitative investigation of the situation above we propose to study the following measure:\newline
\begin{equation}
\label{inter_info}
   I(S:R:C)_{\Omega}:= I(S:R)_{\Omega} - I(S:R|C)_{\Omega},
\end{equation}where in the r.h.s. we have the mutual information and the conditional mutual information for the states in Eqs. (\ref{omega2}) and (\ref{omega1}), respectively. Then, the l.h.s. reveals the difference between the shared information - correlations by quantum systems $\{S,\,R\}$ ($\{R,\,S\}$) when the classical system $C$ has been and has not been considered.

\subsection{Correlations due mutual asymmetry on S+R\label{correlations}}

From now on we will work in the case which $\rho_{SR} = \rho_{S}\otimes \rho_{R}$. The reason that is to keep clear what is the quantum system $R$ that $S$ is reference to. Provided these considerations, we are ready for our first result, relating correlations and asymmetric properties inside a composed quantum system. The proof can be seen in the appendix \ref{appC}.\newline

\begin{lemma} Under the conditions discussed above, the Eq.\ref{inter_info} turns out to be: 
\begin{equation}
I(S:R:C) \equiv \mathcal{A}_{G}(S:R),
\label{oper_mean}
\end{equation}where the r.h.s. is given by the following measure,
\begin{equation}
\mathcal{A}_{G}(S:R) \coloneqq A_{G}(\rho_{S}) + A_{G}(\rho_{R}) - A_{G}(\rho_{SR}),
\label{def1}
\end{equation}which we will call \textit{mutual asymmetry} and $A_{G}(\cdot)$ is given by Eq. (\ref{eq:ent_assym}).\newline
\end{lemma}The mutual asymmetry can be understood as the quantification of the amount of correlations between quantum systems $S$ and $R$ deleting any information residing in a classical reference frame $C$, under global symmetry $G$. In other words, it guarantees that we are only quantifying properties of a quantum system in relation to another one. An analogous of this measure was introduced by the first time at Ref.\cite{jacobs}. Now, given the state $\rho\in\mathcal{H}\equiv\mathcal{H}_{S}\otimes\mathcal{H}_{R}$ as discussed above and the group $G$ with unitary representation $U_{g}^{SR}$ in $\mathcal{H}$, the symmetry in composed systems acts into two ways: \emph{globally} or \emph{locally}. For global symmetry, we have:
\begin{equation}
\mathcal{G}_{G}(\rho_{SR}) = \intop_{g\in G}d\mu(g)\,\mathcal{U}_{g}^{SR}(\rho_{SR})
\label{eq:g_global},
\end{equation}on the other hand, for local symmetry,

\begin{equation}
\mathcal{G}_{G\otimes G}(\rho_{SR})=\intop_{g\in G}\intop_{g'\in G}d\mu(g)d\mu(g')\,\mathcal{U}_{g,g'}^{SR}(\rho_{SR})
\label{eq:g_local}
\end{equation}with $\mathcal{U}_{g,g'}^{SR}\equiv U_{g}^{S}\otimes U_{g'}^{R}(\cdot)U_{g}^{S \dagger}\otimes U_{g'}^{R \dagger}$. The symbol $\mathcal{G}_{G\otimes G}$ indicates that the uniform average acts locally in $S$ and $R$. This splitting it is useful for the fact below: \newline

\begin{prop}\label{my_prop}
Manipulating the expression \ref{oper_mean}, we have that for any compact Lie Group $G$:

\begin{eqnarray}
\mathcal{A}_{G}(S:R) & = & S(\mathcal{G}_{G\otimes G}(\rho_{SR})) - S(\mathcal{G}_{G}(\rho_{SR}))\nonumber\\
& = & S(\mathcal{G}_{G}(\rho_{SR})||\mathcal{G}_{G\otimes G}(\rho_{SR})),
\label{prime_eq}
\end{eqnarray}where $S(\rho||\sigma)=\mbox{tr}\rho(log\rho-log\sigma),\,\forall\rho,\sigma\in\mathcal{H}$. 
\end{prop}This is our prime result, which means that the mutual asymmetry quantifies the difference between one imposes global and local asymmetries in composed systems. It is also important to proves the lower bound of the following lemma. The upper bound is proved in Ref.\cite{jacobs} for some finite and discrete group $G$. In the appendix \ref{appD} we gave a proof for any compact Lie groups promoting shifts in one dimension.

\begin{lemma} The mutual asymmetry satisfies the followings bounds:
\begin{equation}
0\leq\mathcal{A}_{G}(S:R)\leq\min\{A_{G}(\rho_{S}),\,A_{G}(\rho_{R})\},
\end{equation}\label{mutual}with $\rho_{S}$ and $\rho_{R}$ under the same symmetry imposed by the group $G$. The equality is satisfied when given $\rho_{S},$ $\exists\,\rho_{R}$ such that
$\mathcal{A}_{G}(S:R)=A_{G}(\rho_{R})$. This implies that the mutual asymmetry can be seen as a generalization of the relative
entropy of asymmetry.\newline
\end{lemma}

The result above shows that if $A_{G}(\rho_{S})=0$ or $A_{G}(\rho_{R})=0$ implies that $\mathcal{A}_{G}(S:R)=0$, in other words, if either system or reference state is locally symmetric, the mutual asymmetry vanishes. This implies directly the proposition below followed by a mathematical criterion to investigate quantum reference frames inside a globally-symmetric composed systems:\newline

\begin{prop}
Mutual asymmetry between both parts is a necessary condition to have quantum reference frames inside a globally-symmetric composed system.\newline
\end{prop}

\begin{defi}
Let $S+R$ be a composed system under global symmetry imposed by a group $G$. A pair of states $(\rho_{S},\,\rho_{R})$ acts as quantum reference frame for each other iff $\mathcal{A}_{G}(S:R)\neq0$.\newline
\end{defi}

In the forthcoming results motivated by the PWC model we apply the formalism of mutual asymmetry for the case of time-translations group, where the concept of mutual asymmetry turns to be \textit{mutual coherence}. By identifying phase references as clocks, we focus on shifts in one dimension given by $G = U(1)$.\newline

We start by elucidating that a unitary representation of a locally compact Lie group on a Hilbert space $H$ consists of a number of nonequivalent representations called 'charge sectors' $k$ \cite{bartlett}. The Hilbert space can be decomposed into a direct sum of these charge sectors \footnote{In the case of time-translational symmetry the charge sectors turns to be eigenspaces of the Hamiltonian.}, $\mathcal{H}=\bigoplus_{k}\mathcal{H}_{k}$, $\mathcal{H_{S}}=\bigoplus_{m}\mathcal{H}_{m}$ and $\mathcal{H_{R}}=\bigoplus_{n}\mathcal{H}_{n}$, the global symmetry has the following mathematical representation:
\begin{eqnarray}
\mathcal{G}_{G}(\rho_{SR}) & = & \sum_{k}\Pi_{k}\,\rho_{SR}\,\Pi_{k}\nonumber \\
& := & \Pi_{G}(\rho_{SR}),
\label{eq:d_global}
\end{eqnarray}in which $\Pi_{G}(\cdot)$ represents the dephasing map relative to the total Hamiltonian and $U_{g}^{SR}\mathcal{H}_{k}\subset\mathcal{H}_{k}$ are invariant subspaces with $\Pi_{k} \equiv \sum_{m+n=k}\Pi_{m}^{S}\otimes\Pi_{n}^{R}$ the projector onto $\mathcal{H}_{k}$. For the local symmetry representation, we have:
\begin{eqnarray}
\mathcal{G}_{G\otimes G}(\rho_{SR})& = &\sum_{m,n}(\Pi_{m}^{S}\otimes\Pi_{n}^{R})\,\rho_{SR}\,(\Pi_{m}^{S}\otimes\Pi_{n}^{R})\nonumber \\
& := & \Delta(\rho_{SR}),
\label{eq:d_local}
\end{eqnarray}with $\Delta_{G}(\cdot)$ being the \textit{fully} dephasing map now and $U_{g}^{S}\,\mathcal{H}_{m}\subset\mathcal{H}_{m}$, $U_{g}^{R}\,\mathcal{H}_{n}\subset\mathcal{H}_{n}$ invariant subspaces with $\Pi_{m}^{S}$, $\Pi_{n}^{R}$ the projectors onto $\mathcal{H}_{m}$, $\mathcal{H}_{n}$, respectively.\newline

\begin{customdefi}{1'}
For the case of $G$ be the group of time-translation symmetry, the mutual asymmetry $\mathcal{A}_{G}(S:R)$ turns to be the mutual coherence $\mathcal{C}(S:R)$:
\begin{eqnarray}
\mathcal{C}(S:R)& = & S(\Delta(\rho_{SR})) - S(\Pi(\rho_{SR}))\nonumber\\
& = & S(\Pi(\rho_{SR})||\Delta(\rho_{SR})).\newline
\end{eqnarray}
\end{customdefi}Therefore, the mutual coherence is a quantifier which exhibits the existence of correlations due internal coherence \cite{-}. In other words, the measure above is nonzero only when there is a difference between the process of destroying internal from external coherence in global time-symmetric composed systems. Next, we explore the quantum reference orientation for the PWC model considering different regimes for the clock and system states and its relation with good and poor localization. Hereafter, we will deal only with pure states.

\subsection{\label{examples}Some examples}

\paragraph{The qubit model.}$\rho_{\alpha}=|+\left\rangle \right\langle +|$, $\left.|+\right\rangle =\frac{\left.|0\right\rangle +\left.|1\right\rangle }{\sqrt{2}}$

Consider both system $S$ and quantum clock $R$ in the initial state: $\rho_{\alpha}=|+\left\rangle \right\langle +|$, $\left.|+\right\rangle =\frac{\left.|0\right\rangle +\left.|1\right\rangle }{\sqrt{2}}$ which has asymmetry in relation to $U_{\theta}^{\alpha}=\{e^{i\theta \sigma_{z}^{\alpha}}; \theta \in[0,2\pi]\},\:\alpha=S,\,R$. An outside observer under the global symmetry represented by $U_{\theta}=\{e^{i\theta \sigma_{z}}; \theta \in[0,2\pi]\}$ with $\sigma_{z}=\sigma_{z}^{S}\otimes\mathbb{1}_{R}+\mathbb{1}_{S}\otimes\sigma_{z}^{R}$, will attributes the following state:
\begin{equation}
\mathcal{G}_{G}(\rho_{S}\otimes\rho_{R})=\frac{1}{2\pi}\intop_{0}^{2\pi}d\theta\,U_{\theta}\left(\rho_{S}\otimes\rho_{R}\right)U_{\theta}^{\dagger}
\end{equation}

Note that $A_{G}(\rho_{\alpha})=1$ for $\alpha=S,\,R$ and
$A_{G}(\rho_{SR})=\frac{3}{2}$.
$\mathcal{A}_{G}(S:R)=\frac{1}{2}>0$. This case elucidates qualitatively the existence of quantum reference frames in the Page-Wootters universe of two qubits to describe time \cite{83}.\newline

\paragraph{High reference localization.}

Physically, it is expected that a higher localization (which is achieved by a higher dimension of the quantum reference  Hilbert space) of the reference frame $\mathcal{H_{R}}$ gives a better orientation for the system \cite{marvian4,bartlett,love,love1}.\newline

To show this, let us consider the system $S$ in the asymmetric state $\left.|\psi_{S}\right\rangle = \frac{\left.|0\right\rangle +\left.|1\right\rangle }{\sqrt{2}}$. Consider now, the clock as a qudit with Hamiltonian $H_{R}=\sum_{m=0}^{d-1}m|m\left\rangle \right\langle m|$, in which $J_{z}\left.|m\right\rangle =m\left.|m\right\rangle$ and the clock state $R$ in the uniform superposition (maximum likelihood state),

\begin{equation}
\left.|\psi_{R}\right\rangle =\frac{1}{\sqrt{d}}\sum_{m=0}^{d-1}\left.|m\right\rangle,
\end{equation}denoting $\rho_{S}=|\psi_{S}\left\rangle \right\langle \psi_{S}|$ and
$\rho_{R}=|\psi_{R}\left\rangle \right\langle \psi_{R}|$, it is easy to see that $A_{G}(\rho_{S})=1$ e $A_{G}(\rho_{R})=\log d.$\newline

By make some calculations with symmetry imposed now by $U_{\theta}^{\alpha}=\{e^{i\theta J_{z}^{\alpha}}; \theta \in[0,2\pi]\},\:\alpha=S,\,R$ it follows that $A_{G}(\rho_{SR})=\log d+\frac{1}{d}$. Therefore, the mutual asymmetry is
\begin{equation}
\mathcal{A}_{G}(S:R)=1-\frac{1}{d}>0.
\label{opt_qc}
\end{equation}

Note that $\max \mathcal{A}_{G}(S:R)=1$ for $d\rightarrow\infty$. This result implies that increasing the dimension of the clock system $R$ the orientation of $S$ is optimized, in agreement with previous results in the literature.\newline

\begin{figure*}[ht]
\begin{center} 
    \subfloat{{\includegraphics[width=8cm,height=6cm]{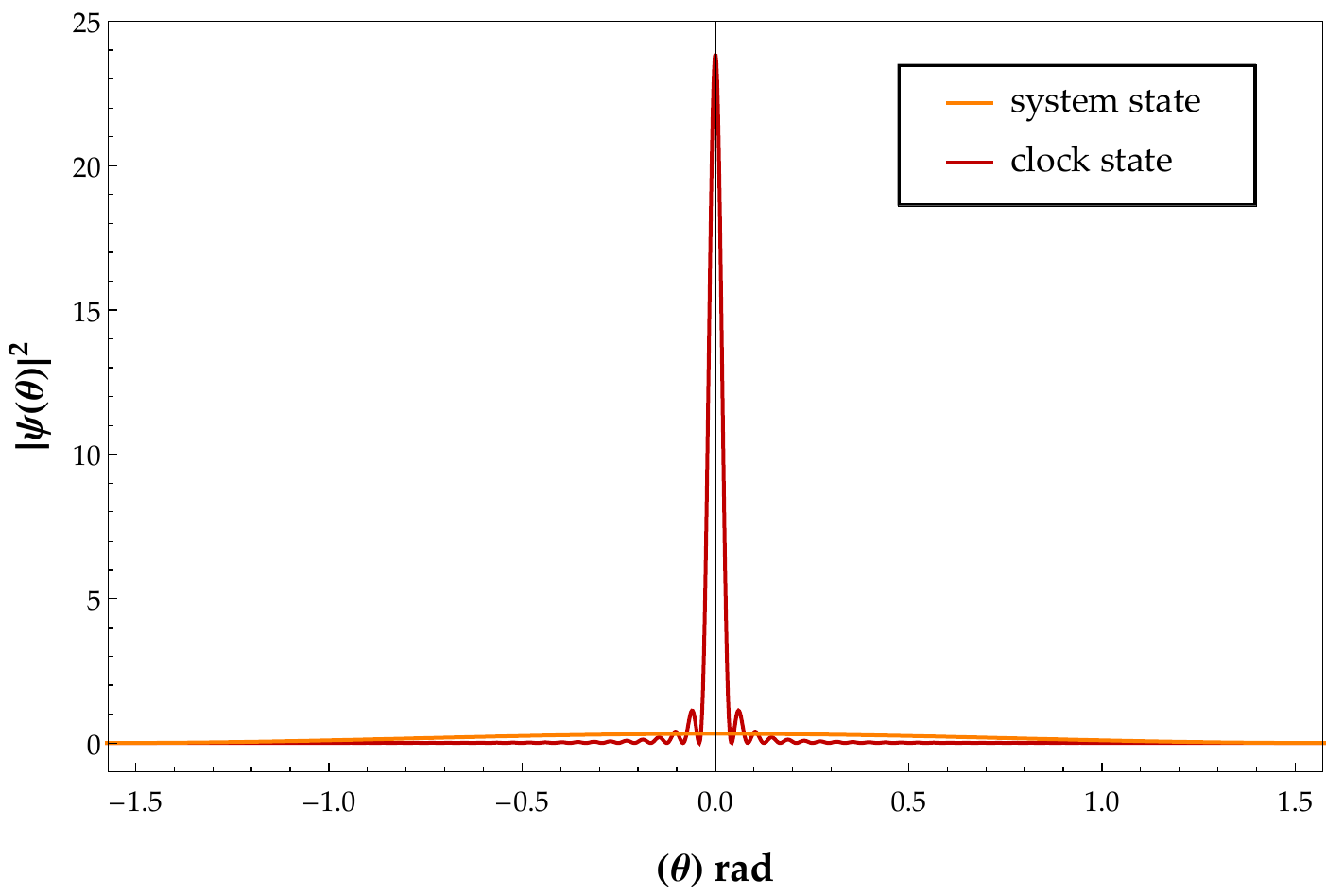}}}%
    \qquad
    \subfloat{{\includegraphics[width=8cm,height=6cm]{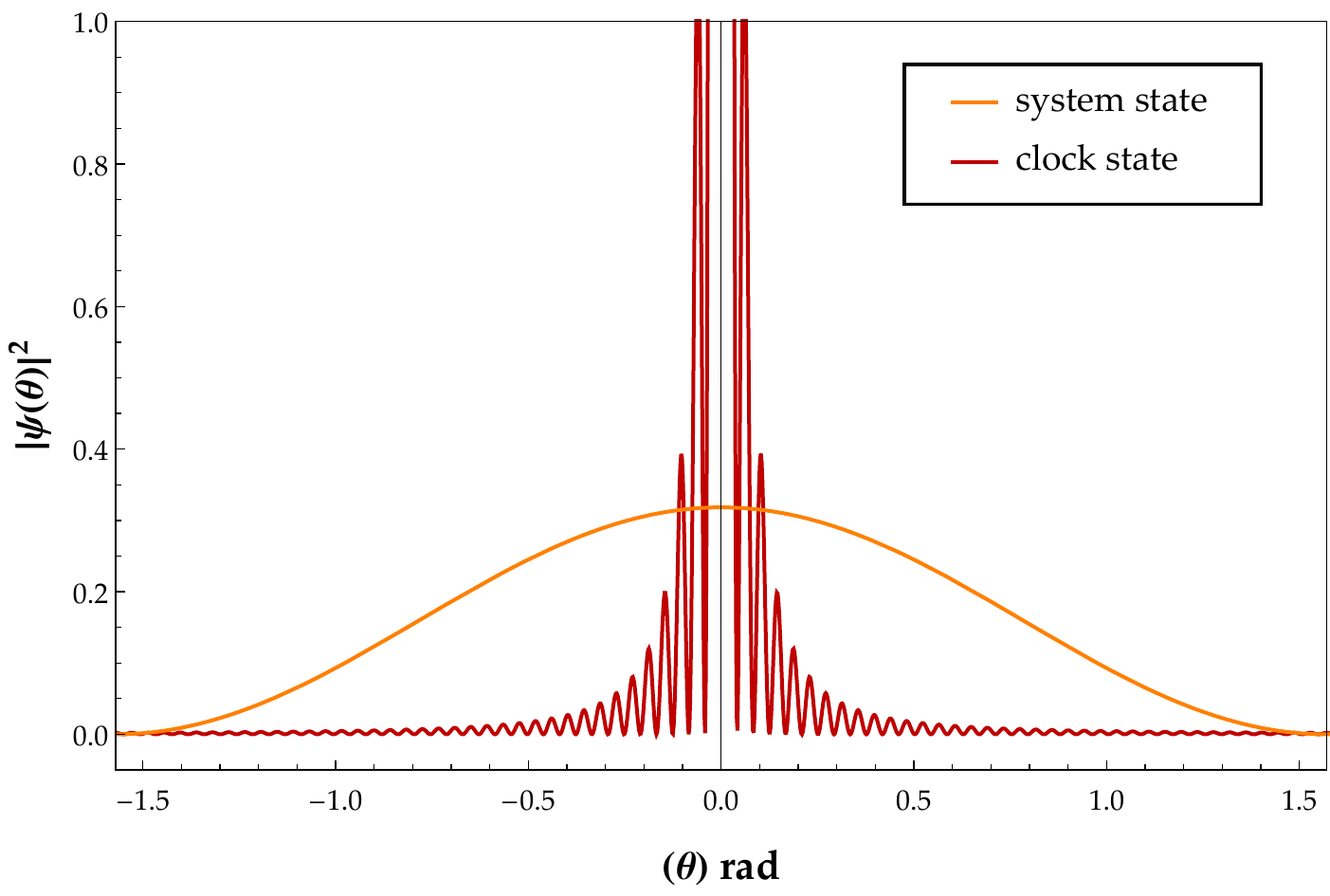}}}%
\end{center}   

\caption{[color online]: Plot for $\psi_{S}(\theta)=\braket{\theta|\psi_{S}}$ and $\psi_{R}(\theta)=\braket{\theta|\psi_{R}}$, in which $\left.|\psi_{S}\right\rangle = \frac{\left.|0\right\rangle +\left.|1\right\rangle }{\sqrt{2}}$ and $\left.|\psi_{R}\right\rangle =\frac{1}{\sqrt{d}}\sum_{m=0}^{d-1}\left.|m\right\rangle,$ $0 \leq \theta<2\pi$. It was considered $d = 150$, by the right-hand zoom the clock wave-function has a narrow peak at $\theta = 0$ and can be viewed as pointing to the $0$-hour with an uncertainty of $\pm \pi/d.$}
\label{fig:image1}
\end{figure*}

\paragraph{High coherence order.}
Given that a system is quantizied along the $z$-axis and $J_{z}\left.|m\right\rangle =m\left.|m\right\rangle$, as the example here, coherence of order $k$ of the state $\rho=\sum_{m,n}\rho_{mn}|m\left\rangle \right\langle n|$ is defined as the 1-norm of the sum of the off-diagonal terms  with $m-n = k$, \cite{marvian1}. Therefore, keeping the clock system as a qudit and considering $S$ in the asymmetric state $\left.|\psi_{S}\right\rangle =\frac{\left.|0\right\rangle +\left.|d-1\right\rangle }{\sqrt{2}}$, we have a state which exhibits coherence of order $d-1$. In this case,
\begin{equation}
\mathcal{A}_{G}(S:R)=\frac{1}{d}>0.
\end{equation}Therefore, even optimizing the clock, which means high dimension and high localization, it is impossible to evaluate a higher order (the same of the clock) of coherence of $S$ using this quantum clock $R$.  Therefore, these last examples clarify how the concepts of local time translation asymmetry and relative coherence coincides. The Figs. \ref{fig:image1}, \ref{fig:image2} clear these facts. To do this, the systems $S$ and $R$ was represented in the angle space, see appendix \ref{appB}, giving the visual aspect of wave function of the clock and system.\newline

\begin{figure*}[ht]

\centering
    \subfloat{{\includegraphics[width=8cm,height=6cm]{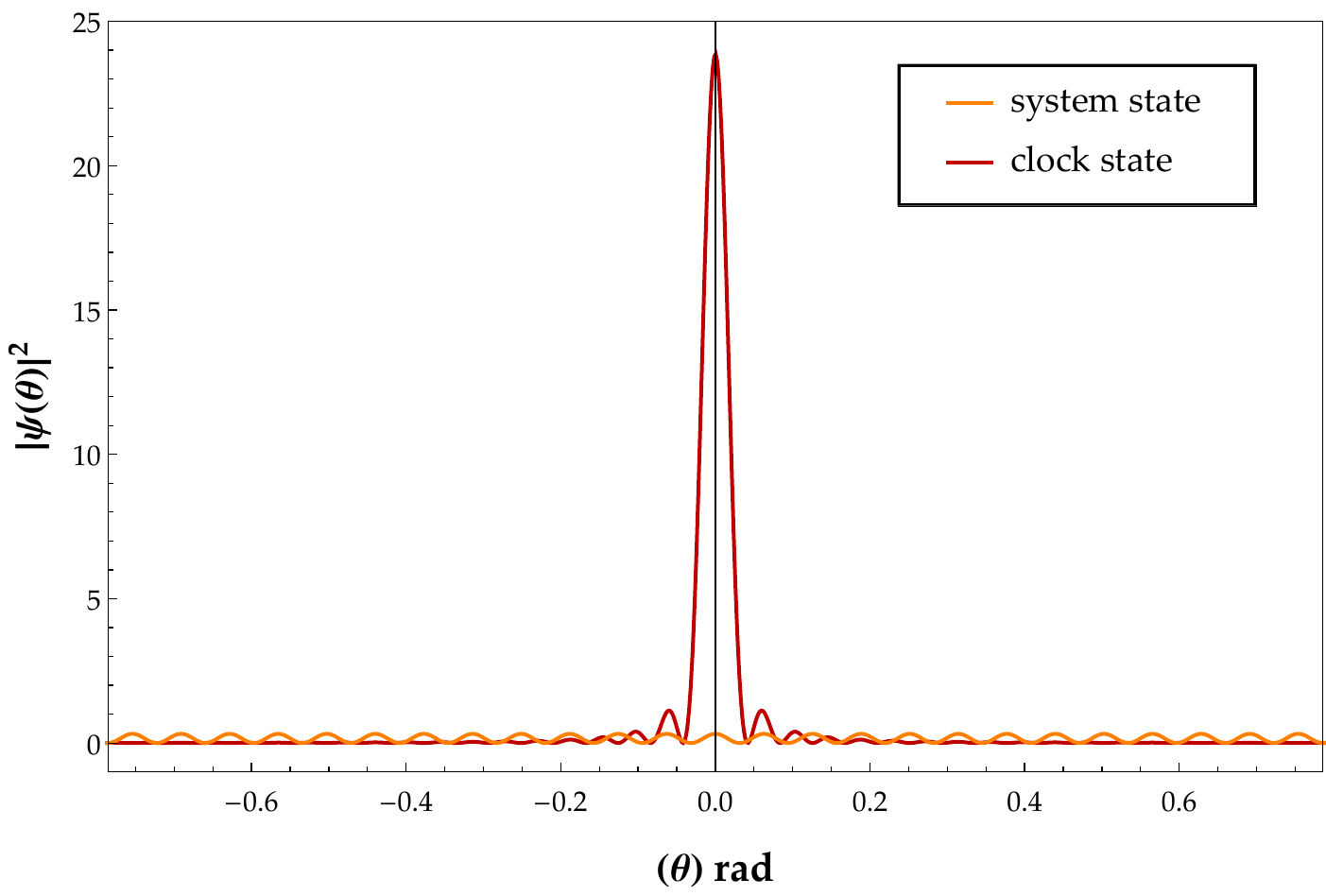}}}%
    \qquad
    \subfloat{{\includegraphics[width=8cm,height=6cm]{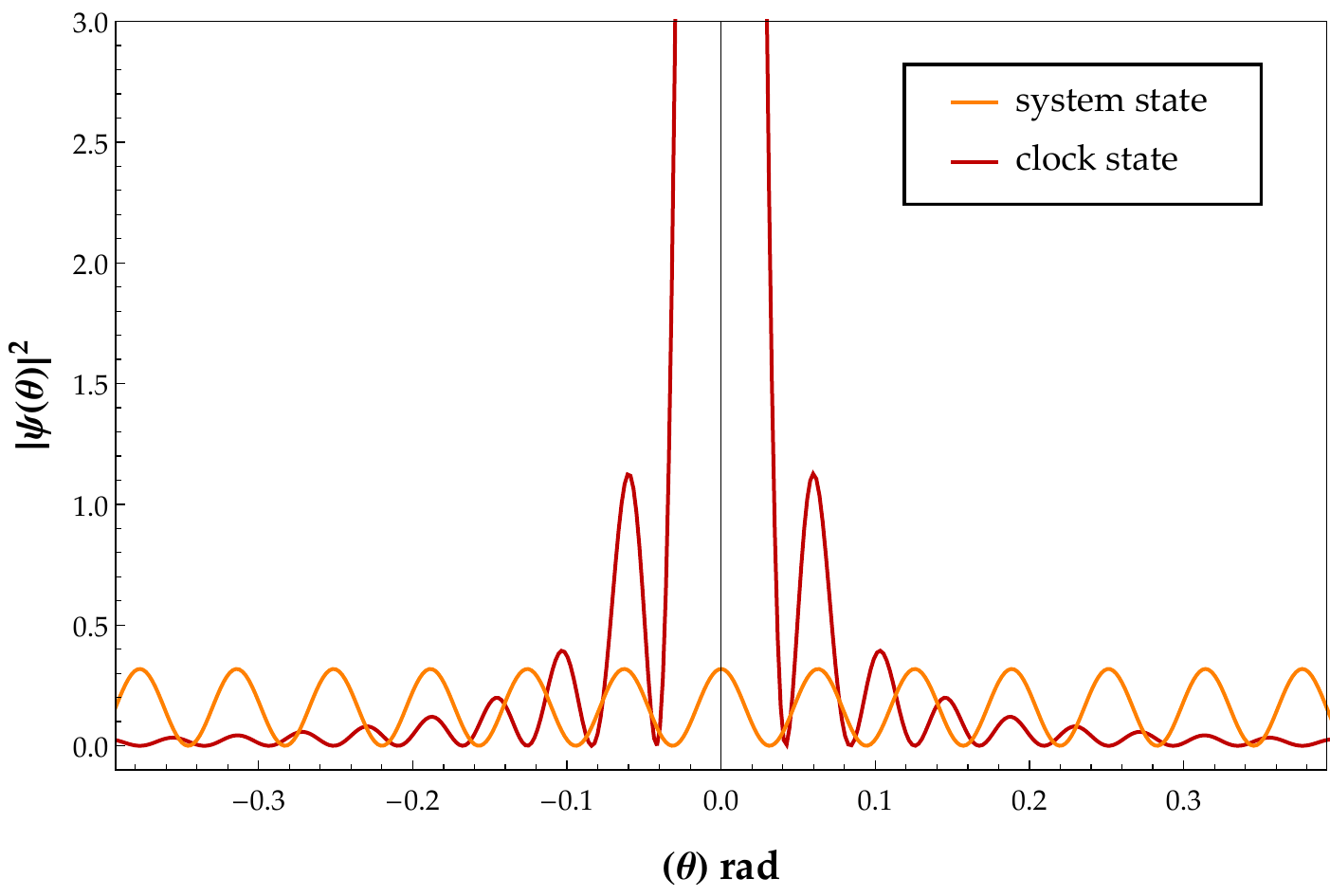}}}%
 
\caption{[color online]: Plot for $\psi_{S}(\theta)=\braket{\theta|\psi_{S}}$ and $\psi_{R}(\theta)=\braket{\theta|\psi_{R}}$, in which $\left.|\psi_{S}\right\rangle = \frac{\left.|0\right\rangle +\left.|d-1\right\rangle }{\sqrt{2}}$ and $\left.|\psi_{R}\right\rangle=\frac{1}{\sqrt{d}}\sum_{m=0}^{d-1}\left.|m\right\rangle,$ $0 \leq \theta<2\pi$. It was considered $d = 150$, here the narrow peak of the clock state is not sufficient to localize the high coherence order of the system state which can be confirmed by the small variance of $\psi_{S}(\theta)$ around $\theta = 0$, as we can see by the right-hand zoom.}
\label{fig:image2}

\end{figure*}


We can verify the role of internal coherence by the analytic expression for $\mathcal{G}_{G}(\rho_{SR})$ in the three examples worked previously, which are given below, respectively. The off-diagonal terms, dashedbox in the expressions, those that provide the observation of internal coherence are responsible by correlations in the globally-symmetric density operator. Furthermore, note that, Eq. (\ref{degenerescence}) confirms that high degenerescence of the null eigenvalue of the total Hamiltonian, gives a better internal quantum clock, by Eq. (\ref{opt_qc}). This can be clarified by using Eq. (\ref{first}). This implies that, the density operator and total Hamiltonian has the same eigenbasis and diagonalizing Eq. (\ref{degenerescence}) give us the result.\newline

\begin{widetext}
\begin{equation}
\frac{1}{2}\Bigg[\sum_{n=0}^{1}\Big(\ket{n,0}\bra{n,0}+\ket{n,1}\bra{n,1}\Big)+\dashbox{$\ket{0,1}\bra{1,0}+\ket{1,0}\bra{0,1}$}\Bigg],
\end{equation}

\begin{equation}
\frac{1}{2d}\sum_{n=0}^{d-1}\Bigg(|n,0\left\rangle \right\langle n,0|+|n,1\left\rangle \right\langle n,1|+\dashbox{$\ket{n+1,0}\bra{n,1}+|n,1\left\rangle \right\langle n+1,0|$}\Bigg),
\label{degenerescence}
\end{equation}

\begin{equation}
\frac{1}{2d}\Bigg[\sum_{n=0}^{d-1}\Big(|n,0\left\rangle \right\langle n,0|+|n,d-1\left\rangle \right\langle n,d-1|\Big)+\dashbox{$\ket{0,d-1}\bra{d-1,0} +\ket{d-1,0}\bra{0,d-1}$}\Bigg].
\end{equation}
\end{widetext}

\section{\label{conc}Conclusion}

In this work, we introduced an operational formulation to investigate quantum reference frames inside composed systems. To do so, we proposed to evaluate the shared correlation between a symmetric composed system removing all existing classical reference information. This led us to a quantifier named mutual asymmetry, which we gave an operational interpretation as well as it importance to identify quantum reference frames by splitting the concept of symmetries into global and local.\newline

Provided that, we were able to give a conceptually clear formulation of the Page-Wootters clock (PWC). By modeling time-symmetry using the action of U(1)-phase group and applying the formulation of mutual asymmetry, which turns to be mutual coherence in this particular case, we verified the importance of internal coherence in the PWC. It was verified, by using the quantifier, that greater the dimension of the clock reference space better the time-orientation. Also, the difference between the dimension of system and clock reference spaces has to be of a moderate value to have a reasonable quantum reference frame in high dimensions.\newline

We hope that the discussions made here helps in some way the formulation of quantum concepts without a classical background to reference to, i.e., considering the whole description - system \textit{plus} reference inside quantum theory. Indeed, our approach has physical similarities with that developed by Loveridge et al.\cite{love1}, in respect to an internal character of the quantum theory.\newline

Finally, we raise some questions. The general character of the mutual asymmetry $\mathcal{A}(S:R)$ could allow to explore it in the context of gauge symmetries. Indeed, in Ref.\cite{jennings} is investigated an information-theoretic analysis of gauging a global symmetry to a local one in terms of reference frames. By doing the gauge at the level of states, $\mathcal{A}(S:R)$ can be seen as a measure of the correlations among the systems with the gauge fields. Beyond that, the measure given by Eq. (\ref{inter_info}) was already discussed in the classical information theory. It is the special three-variable case of the named \textit{interaction information} which can assume negative values, \cite{fano}. Here, this happens when one works outside the regime of product states. This could open the door to investigate the physical meaning to deal with quantum reference frames inside composed systems with unknown states initially. \newline

\section{\label{agrad}Acknowledgments}

The authors would like to Leandro R. S. Mendes for reading the paper carefully and his valuable comments. We also thank you to David Jennings to clarify the application of our approach in gauge symmetries. The project was funded by Brazilian funding agencies CNPq (Grants No.140665/2018-8, 305201/2016-6), FAPESP (Grant No.2017/03727-0, 2017/07973-5) and the Brazilian National Institute of Science and Technology of Quantum Information (INCT/IQ).


\clearpage

\widetext
\appendix

\begin{center}
\textbf{\Large Supplementary Material}
\end{center}

\setcounter{equation}{0}
\setcounter{figure}{0}
\setcounter{table}{0}

\makeatletter
\renewcommand{\theequation}{S\color{blue}\arabic{equation}}
\renewcommand{\thefigure}{S\arabic{figure}}
\renewcommand{\bibnumfmt}[1]{[S#1]}

\vspace{0.5cm}
\begin{center}
\section{\label{appA}\normalsize The internal dynamics and measurement in PWC model}
\end{center}
\vspace{0.5cm}

In the special case that the clock does not interact with the system as already mentioned we have that:

\begin{equation}
H = H_{S}\otimes \mathbb{1}_{R}+\mathbb{1}_{S}\otimes H_{R}.
\label{s1}
\end{equation}

The correlations of the following time-symmetric vector in the composed system,
\begin{equation}
\left.\left.|\psi\right\rangle \right\rangle =\sum_{t}c_{t}\left.|\psi_{S}(t)\right\rangle \left.|\phi_{R}(t)\right\rangle,
\end{equation}gives the internal dynamics in the system $S$. The system history is given by a sequence of events codified in the various $\left.|\psi_{S}(1)\right\rangle ,\left.|\psi_{S}(2)\right\rangle,\ldots,\left.|\psi_{S}(t)\right\rangle$, each describing the state of $S$ with respect to the clock $R$ given that the latter is in the state $\ket{\phi_{R}(t)}$ at clock time $t$ \cite{16}. The internal dynamics follows the Schr{\"o}dinger equation with respect to the parameter $t$:
\begin{equation}
\frac{d\rho_{S}(t)}{dt}=i\left[\rho_{S}(t),H_{S}\right],\,\,\rho_{S}(t) = \ket{\psi_{S}(t)}\bra{\psi_{S}(t)}.\newline
\end{equation}

To see this, consider $\left.|\phi_{R}(0)\right\rangle $ as the zero hour of the clock and for each $t$ we have that $\left.|\phi_{R}(t)\right\rangle =\exp\left(-iH_{R}t\right)\left.|\phi_{R}(0)\right\rangle $. Let $\left.|\psi_{S}(t)\right\rangle $ be the relative state of system $S$ when the clock system $R$ is in the state $\left.|\psi_{R}(t)\right\rangle $.
In other words, $\left.|\psi_{S}(t)\right\rangle $ is the result of the projection \footnote{This projection has not to do with a measurement process.} $\left.\left.|\psi\right\rangle \right\rangle $ in the clock system subspace,
\begin{equation}
\left.|\psi_{S}(t)\right\rangle =\left\langle \phi_{R}(t)\left.|\psi\right\rangle \right\rangle.
\end{equation}

Now, using the fact that $H\left.\left.|\psi\right\rangle \right\rangle=0$ and Eq.(\ref{s1}) in the clock representation $\mathcal{H}_{R}$:

\begin{eqnarray}
i\frac{\partial}{\partial t}\left.|\psi_{S}(t)\right\rangle  & = & i\frac{\partial}{\partial t}\left\langle \phi_{R}(t)\left.|\psi\right\rangle \right\rangle =-\left\langle \phi_{R}(t)\left.|H_{R}\otimes I_{S}|\psi\right\rangle \right\rangle \nonumber \\
 & = & -\left\langle \phi_{R}(t)\left.|H_{tot}-H_{S}\otimes I_{R}|\psi\right\rangle \right\rangle \nonumber \\
 & = & \left\langle \phi_{R}(t)\left.|H_{S}\otimes I_{R}|\psi\right\rangle \right\rangle \nonumber \\
 & = & H_{S}\left\langle \phi_{R}(t)\left.|\psi\right\rangle \right\rangle =H_{S}\left.|\psi_{S}(t)\right\rangle. \nonumber 
\end{eqnarray}

For a density matrix $\rho_{S}(t)=\sum_{j}p_{j}\ket{\psi_{S,j}(t)}\bra{\psi_{S,j}(t)}$ we can note that, considering for only one term $j$ and the others are similar,\newline

\begin{eqnarray}
\frac{\partial \rho_{S}(t)}{\partial t} & = & \frac{\partial}{\partial t}\Big(\ket{\psi_{S}(t)}\bra{\psi_{S}(t)}\Big) = \frac{\partial}{\partial t}\Big(\ket{\psi_{S}(t)} \Big)\bra{\psi_{S}(t)}+\ket{\psi_{S}(t)}\frac{\partial}{\partial t}\Big(\bra{\psi_{S}(t)}\Big) \nonumber \\
& = & -i H_{S}\ket{\psi_{S}(t)}\bra{\psi_{S}(t)}+i \ket{\psi_{S}(t)}\bra{\psi_{S}(t)}H_{S}\nonumber \\
& = & i\left[H_{S},\rho_{S}(t)\right], \nonumber
\end{eqnarray}evolving according to the Liouville-von Neumann equation with respect to the clock time $t$.\newline

Even considering a time-independent Hamiltonian in the above case, this construction is also compatible with a time-dependent Hamiltonian arising on a subsystem of the system $S$. The time-dependent Hamiltonian for this subsystem can be seen as an approximate description due the interactions between the subsystem and the environment, \cite{16, briggs, 93}.\newline

In this mechanism, measurements of a physical quantity in the system $S$ at a given clock time for the clock $R$ are described by the conditional probability formalism \cite{15,monte}. To elucidate, assume that the time-symmetric composed state is described by the density matrix $\rho$. Then, the conditional probability to obtain the eigenvalue $o$ for the quantity $O_{S}\in \mathcal{H}_{S}$ given the eigenvalue $t$ for $T_{R} \in \mathcal{H}_{R}$ is:

\begin{equation}
p(o|t)_{\rho}=\lim_{\tau\rightarrow\infty}\frac{\intop_{-\tau}^{\tau}dT\,[P_{o}(T)P_{t}(T)\rho P_{t}(T)]}{\intop_{-\tau}^{\tau}dT\,[P_{t}(T)\rho]},
\end{equation}where the quantity $P_{o}(T)$ is the projector onto the eigenspace associated with the  eigenvalue $o$ of the operator $O_{S}$ at coordinate
time $T$ and similarly for $P_{t}(T)$. Notice that the expression does not require assigning a value to the classical parameter $T$, since it is integrated over all possible value. A generalization of this expression to multiple time measurements can be seen in Ref.\cite{monte1}.

\vspace{0.5cm}
\section{\normalsize A finite cyclic quantum clock model\label{appB}}
\vspace{0.5cm}

The Peres-Salecker-Wigner clock \cite{80} gives the quantum clock as a qudit. The Hamiltonian of the system is given by,  
\begin{equation}
H_{R}=\sum_{m=0}^{d-1}\frac{2m\pi}{d} |m\left\rangle \right\langle m|,
\end{equation}with an orthogonal eigenbasis $\{{\left.|m\right\rangle}\}_{m=0}^{d-1}$, $\left\langle m|m'\right\rangle=\delta_{m,m'}$. The generator can be seen as:

\begin{equation}
U_{R}(m)=\sum_{m=0}^{d-1} e^{-i 2m\pi/d} |m\left\rangle \right\langle m|,  
\end{equation}noting that $U_{R}(m)=(U_{R}(1))^{m}$, $\forall m\in \mathbb{N}$ and $U_{R}(d)=U_{R}(0)=\mathbb{1}_{R}$. The clock operator can be defined as:
\begin{equation}
T_{R}=\sum_{k=0}^{d-1}k|k\left\rangle \right\langle k|,
\end{equation}where the eigenbasis $\{{\left.|k\right\rangle}\}_{k=0}^{d-1}$, with $\left\langle k|k'\right\rangle=\delta_{k,k'}$, is given by the discrete Fourier transform of the states $\{{\left.|m\right\rangle}\}_{m=0}^{d-1}$,
\begin{equation}
\left.|k\right\rangle=\frac{1}{\sqrt{d}}\sum_{m=0}^{d-1}e^{-i2m\pi k/d}\left.|m\right\rangle.
\end{equation} 

It is interesting to note that $U_{R}(1)\left.|k\right\rangle=\left.|k+1\right\rangle$ and $U_{R}(1)\left.|d-1\right\rangle=\left.|0\right\rangle$. Consider now, the previous states in the angle representation $0 \leq \theta<2\pi$,
\begin{equation}
m(\theta)=\braket{\theta|m}=\frac{1}{\sqrt{2\pi}}e^{im\theta},
\end{equation}
\begin{eqnarray}
k(\theta)=\braket{\theta|k}  & = & \frac{1}{\sqrt{d}}\sum_{m=0}^{d-1}e^{-2\pi ikm}m(\theta) \protect \nonumber \\
& = & \frac{1}{\sqrt{2\pi d}}\sin\left[\frac{d}{2}\left(\theta-\frac{2\pi k}{d}\right)\right]\Bigg/\sin\left[\frac{1}{2}\left(\theta-\frac{2\pi k}{d}\right)\right]
\end{eqnarray}
\vspace{0.5cm}

The greater the values of $d$, narrower is the peak of these functions in $\theta=\frac{2\pi k}{d}$ and they can be visualized as pointing to the $k-th$ hour with uncertainty $\pm \pi/d$ \cite{80}.\newline

However, this Hamiltonian and clock operator does not satisfies $\bra{k}[T_{R},\,H_{R}]\ket{k}=0,\,\forall k$ due their discrete character \cite{80}. Indeed, as already observed by Weyl \cite{pashby}, the canonical commutation relation cannot be satisfied for finite dimensional operators.
To overcome this, using this model yet, it is possible $T_{R}$ and $H_{R}$ achieve the canonical commutation relation after one restricts the domain of these operators to a sub-domain. This construction is present in the literature under the name of gaussian clock states \cite{qc3}. The sub-domain consists of gaussian superposition of the clock states $\{\ket{k}\}_{k=0}^{d-1}$ excluding pure angle states. Then, $[T_{R},\,H_{R}]\ket{\Psi}\approx i\ket{\Psi}$
for the new clock states $\ket{\Psi}$, where the approximation becomes exact when $d \rightarrow \infty$. The interesting for us is the fact that their initial state - zero hour coincides with that worked by us in the main text.  
\vspace{0.5cm}`

\begin{center}
\section{\normalsize Proof of the relation \texorpdfstring{\ref{oper_mean}}.\label{appC}}
\end{center}
\vspace{0.5cm}


\begin{proof}We will make use of the lemma \ref{design} to help write the group averaging operation for a compact Lie group by a sum of discrete distributions of the symmetry imposed in the state $\Omega_{CSR}$:
\begin{equation}
\Omega_{CSR}=\intop_{g\in G}d\mu(g)\,\ket{g}\bra{g}^{C}\otimes\mathcal{U}_{g}^{SR}(\rho_{SR})=\sum_{i}p_{i}\,\ket{g_{i}}\bra{g_{i}}^{C}\otimes\mathcal{U}_{g_{i}}^{SR}(\rho_{SR}),
\end{equation}with $K=\{g_{i}\}_{i=1}^{m(d)}\subset G$ and $\{p_{i}\}_{i=1}^{m(d)}\in \mathbb{R}$ weighting probabilities. And, to keep the  condition of classical reference frame for $C$, $\{\ket{g_{i}} ;\, g_{i} \in G\}$ is also an orthogonal set of states.\newline

Now, we start denoting $\Omega_{S}=\mbox{tr}_{R}\Omega_{SR}$ and $\Omega_{R}=\mbox{tr}_{S}\Omega_{SR}$. Then, when we calculate the conditional mutual information,

\begin{equation}
    I(S:R|C)_{\Omega} = S(\Omega_{SC}) + S(\Omega_{RC}) - S(\Omega_{SRC}) - S(\Omega_{C}),
    \label{cmi_1}
\end{equation}we have that:

\begin{equation}
    I(S:R|C)_{\Omega} = S(\rho_{S}) + S(\rho_{R}) - S(\rho_{SR}),
    \label{cmi_2}
\end{equation}that is, the mutual information between $\rho_{S}$ and $\rho_{R}$. From Eq.(\ref{cmi_1}) to Eq.(\ref{cmi_2}) we use the joint entropy theorem \cite{nielsen} as follows: 

\begin{eqnarray}
S(\Omega_{CSR}) &=& S\big(\sum_{i}p_{i}\,\ket{g_{i}}\bra{g_{i}}^{C}\otimes\mathcal{U}_{g_{i}}^{SR}(\rho_{SR})\big)\nonumber\\
 &=& H(p_{i}) + \sum_{i}p_{i}S\big(\mathcal{U}_{g_{i}}^{SR}(\rho_{SR})\big)\nonumber\\
&=& H(p_{i}) + S(\rho_{SR}),
\end{eqnarray}in which we use the three facts: the ortonormality of the set $\{\ket{g_{i}} ;\, g_{i} \in G\}$, the invariance of von-Neumann entropy under unitary transformations and $\sum_{i}p_{i}=1$. The argument is similar to calculate $S(\Omega_{CS})$ and $S(\Omega_{CR})$. The term $H(p_{i})$ appeals in the four expressions, however, they cancel due the conditional mutual information structure. Finally, the mutual information,
\begin{equation}
    I(S:R)_{\Omega} = S(\Omega_{S}) + S(\Omega_{R}) - S(\Omega_{SR}),
\end{equation}turns out to be equal to:

\begin{equation}
    I(S:R)_{\Omega} = S(\mathcal{G}_{G}(\rho_{S})) + S(\mathcal{G}_{G}(\rho_{R})) - S(\mathcal{G}_{G}(\rho_{SR})),
\end{equation} in this way the relation \ref{oper_mean} in the main text follows straightforward.
\end{proof}

\begin{lemma}[\cite{approx_1,approx_2}]\label{design} Given a group $G$ with a unitary representation $U$ of dimension $d$, there exists a finite set $K=\{g_{i}\}_{i=1}^{m(d)}\subset G$ and weighting probabilities $\{p_{i}\}_{i=1}^{m(d)}\in \mathbb{R}$, such that:

\begin{equation}
\intop_{g\in G}d\mu(g)\,\mathcal{U}_{g}(\rho)=\sum_{i=1}^{m(d)}p_{i}\,\mathcal{U}_{g_{i}}(\rho),
\end{equation}for all states $\rho$. Here $m(d)$ denotes the number of terms and satisfies the upper bound $m(d)\leq d^{2}$.

\label{lemma_approx}
\end{lemma}

\begin{center}
\section{\label{appD}\normalsize General upper bound on mutual asymmetry (lemma \ref{mutual})}
\end{center}
\vspace{0.5cm}

\begin{proof}
First, remember that $A_{G}(\cdot)$ was defined using Holevo's monotone \cite{marvian1}, in other words,  $A_{G}(\cdot) \equiv \chi \{p_{unif},\,\mathcal{U}_{g}(\cdot)\}$ \footnote{the distribution $p_{unif}$ is the delta distribution at the identity of group.}. Now, if $\chi_{S}$ and $\chi_{R}$ are the Holevo's monotones for the subsystems $S$ and $R$, respectively, using that $\chi$ is non-increasing under partial
trace \cite{96w},

\begin{equation}
A_{G}(\rho_{S}\otimes\rho_{R})\geq A_{G}(\rho_{\alpha}),\:\alpha=S,\,R.
\end{equation}

To show the equality, we choose a normalized state in $R$ on a eigenspace of sufficiently large dimension, in other words, $\rho_{R} \propto \Pi_{n}$ with $n \approx k$, where $\mathcal{G}(\rho_{S} \otimes \rho_{R}) = \sum_{k}\Pi_{k}(\rho_{S} \otimes \rho_{R})\Pi_{k}$\,, $\Pi_{k} = \sum_{m+n=k}\Pi_{m}^{S}\otimes\Pi_{n}^{R}$: 

\begin{eqnarray}
A_{G}(\rho_{S}\otimes\rho_{R}) & \approx & S(\rho_{S}) + S(\mathcal{G}(\rho_{R})) - S(\rho_{S}) - S(\rho_{R})\\ \nonumber
& = & A_{G}(\rho_{R}),
\end{eqnarray}where it was used $\mathcal{G}(\rho_{S} \otimes \rho_{R}) \propto \Pi_{m}\rho_{S}\Pi_{m}\otimes \sum_{n}\Pi_{n}\rho_{R}\Pi_{n}$ and the fact that entropy is additive.\newline

Putting all together,

\begin{equation}
\mathcal{A}_{G}(S:R) = A_{G}(\rho_{S}) + A_{G}(\rho_{R}) - A_{G}(\rho_{R}) = A_{G}(\rho_{S})
\end{equation}
\end{proof}

\begin{center}
\section{\normalsize Proof of proposition \ref{my_prop}\label{appE}}
\end{center}
\vspace{0.5cm}

\begin{proof}
First, note that for $\rho_{SR}=\rho_{S}\otimes \rho_{R}$:
\begin{eqnarray}
\mathcal{A}_{G}(S:R) & = & A_{G}(\rho_{S})+A_{G}(\rho_{R})-A_{G}(\rho_{SR})\nonumber\\
& = & [S(\mathcal{G}_{G}(\rho_{S}))+S(\mathcal{G}_{G}(\rho_{R}))-S(\mathcal{G}_{G}(\rho_{SR}))] - [S(\rho_{S})+S(\rho_{R})-S(\rho_{SR})]\nonumber\\
& = & S(\mathcal{G}_{G}(\rho_{S})\otimes\mathcal{G}_{G}(\rho_{R})) - S(\mathcal{G}_{G}(\rho_{SR}))\nonumber\\
& = & S(\mathcal{G}_{G \otimes G}(\rho_{SR})) - S(\mathcal{G}_{G}(\rho_{SR})).\label{rel_entropy2}
\end{eqnarray}

Now, note that for a general $\rho_{SR}$, the last expression can be written as:\newline

\begin{eqnarray}
S(\mathcal{G}_{G\otimes G}(\rho_{SR}))-S(\mathcal{G}_{G}(\rho_{SR}))
& = & \tr{\mathcal{G}(\rho_{SR})\log\mathcal{G}(\rho_{SR})}-\tr{\mathcal{G}_{G\otimes G}(\rho_{SR})\log\mathcal{G}_{G\otimes G}
(\rho_{SR})}\nonumber \\
& = & \tr{\mathcal{G}(\rho_{SR})\log\mathcal{G}(\rho_{SR})}-
\tr{\mathcal{G}(\rho_{SR})\log\mathcal{G}_{G\otimes G}(\rho_{SR})}\nonumber \\
& = & S(\mathcal{G}_{G}(\rho_{SR})||\mathcal{G}_{G\otimes G}(\rho_{SR})),\nonumber
\end{eqnarray} in which in the second equality it was used that $\mathcal{G}(\rho)$ is $G$-invariant ($\mathcal{G}\circ\mathcal{U}=\mathcal{\mathcal{U}}\circ\mathcal{G}$) $\rightarrow$ $\tr{\rho\log\mathcal{G}(\rho)}=\tr{\mathcal{U}(\rho)\log\mathcal{G}(\rho)},\,\forall g\in G$ and $\intop_{g\in G}d\mu(g)\,\tr{\rho}=\intop_{g\in G}d\mu(g)\,\tr{\mathcal{U}(\rho)}=\tr{\intop_{g\in G}d\mu(g)\,\,\mathcal{U}(\rho)}$.\newline
\end{proof}

\end{document}